\documentclass[preprint]{aastex}  

\usepackage{graphicx}
\usepackage{txfonts}
\usepackage{natbib}
\usepackage{rotating}
\usepackage{float}
\usepackage{lscape}
\usepackage{array}
\usepackage{verbatim}
\usepackage{mathenv}
\usepackage{lscape}

\begin{document}

   \title{Markarian 6: shocking the environment of an intermediate Seyfert}


   \author{B. Mingo \altaffilmark{1}, M. J. Hardcastle \altaffilmark{1}, J. H. Croston \altaffilmark{2}, D. A. Evans \altaffilmark{3}, A. Hota \altaffilmark{4}, P. Kharb \altaffilmark{5}, R. P. Kraft \altaffilmark{6}
          }

   \altaffiltext{1}{School of Physics, Astronomy \& Mathematics, University of Hertfordshire, College Lane, Hatfield AL10 9AB, UK}
		\altaffiltext{2}{School of Physics and Astronomy, University of Southampton, Southampton SO17 1SJ, UK}
		\altaffiltext{3}{Massachusetts Institute of Technology, Kavli Institute for Astrophysics and Space Research, 77 Massachusetts Avenue, Cambridge, MA 02139, USA}
		\altaffiltext{4}{Institute of Astronomy and Astrophysics, Academia Sinica, P.O. Box 23-141, Taipei 10617, Taiwan}
		\altaffiltext{5}{Department of Physics, Rochester Institute of Technology, Rochester, NY 14623, USA}
		\altaffiltext{6}{Harvard-Smithsonian Center for Astrophysics, 60 Garden Street, Cambridge, MA 02138, USA}

   \date{ApJ Received August 25th 2010 ; accepted January 28th 2011}

\begin{abstract}
Markarian 6 is a nearby (D$\sim$78 Mpc) Seyfert 1.5, early-type galaxy, with a double set of radio bubbles. The outer set spans $\sim$7.5 kpc and is expanding into the halo regions of the host galaxy. We present an analysis of our new \textit{Chandra} observation, together with archival \textit{XMM-Newton} data, to look for evidence of emission from shocked gas around the external radio bubbles, both from spatially resolved regions in \textit{Chandra} and from spectral analysis of the \textit{XMM} data. We also look for evidence of a variable absorbing column along our line of sight to Mrk 6, to explain the evident differences seen in the AGN spectra from the various, non-contemporaneous, observations.
We find that the variable absorption hypothesis explains the differences between the \textit{Chandra} and \textit{XMM} spectra, with the \textit{Chandra} spectrum being heavily absorbed. The intrinsic $N_H$ varies from $\sim8\times 10^{21}$ atoms cm$^{-2}$ to $\sim3\times 10^{23}$ atoms cm$^{-2}$ on short timescales (2-6 years). The past evolution of the source suggests this is probably caused by a clump of gas close to the central AGN, passing in front of us at the moment of the observation. Shells of thermal X-ray emission are detected around the radio bubbles, with a temperature of $\sim$0.9 keV. We estimate a temperature of $\sim$0.2 keV for the external medium using luminosity constraints from our \textit{Chandra} image. We analyse these results using the Rankine-Hugoniot shock jump conditions, and obtain a Mach number of $\sim$3.9, compatible with a scenario in which the gas in the shells is inducing a strong shock in the surrounding ISM. This could be the third clear detection of strong shocks produced by a radio-powerful Seyfert galaxy. These results are compatible with previous findings on Centaurus A and NGC 3801, supporting a picture in which these AGN-driven outflows play an important role in the environment and evolution of the host galaxy.
\end{abstract}


   \maketitle

%

\section{Introduction}\label{Intro}

Recent \textit{Chandra} observations of the environments of several powerful radio galaxies (e.g. Hydra A, \citeauthor{McNamara2000} \citeyear{McNamara2000}; M87, \citeauthor{Young2002} \citeyear{Young2002}; Hercules A, \citeauthor{Nulsen2005} \citeyear{Nulsen2005}; see also the review by \citeauthor{McNamara2007} \citeyear{McNamara2007}) have led to significant progress in understanding the AGN-driven gas outflows in these systems and the role they play in galaxy formation and evolution \citep{Croton2006, Bower2006}.

We now know that, although the most powerful radio outflows, spanning hundreds of kpc, are associated with massive elliptical systems, smaller structures also connected to an active nucleus can be found in a variety of systems and environments, including spiral and disk galaxies (see e.g. \citeauthor{Gallimore2006} \citeyear{Gallimore2006}, \citeauthor{Kharb2006} \citeyear{Kharb2006}, \citeauthor{Hota2006} \citeyear{Hota2006}, \citeauthor{Saikia2010} \citeyear{Saikia2010}). The mechanism by which these structures are produced is likely to be related to the one we see in the most powerful sources, but on a smaller scale. Most of the observed AGN-driven bubbles have been found to be overpressured with respect to their surroundings, and may be thus inducing shocks into their surrounding medium. The wide range of morphologies of the galaxies where radio bubbles have been found, and the fact that this AGN-driven phenomenon is most likely episodic \citep{Saikia2010} make understanding the energetics involved in this process fundamental to estimate its impact on AGN feedback and galaxy evolution, and extrapolate how common this mechanism can be in low-power systems. 

In our search for evidence of galaxy feedback associated with kpc-scale radio bubbles, we have carried out observations of a variety of systems, the most notable perhaps being the nearby Fanaroff-Riley type I (FR I, \citeauthor{FR1974} \citeyear{FR1974}) galaxy Centaurus A (\citeauthor{Kraft2003} \citeyear{Kraft2003}, \citeauthor{Croston2009} \citeyear{Croston2009}), finding evidence for shocks also in smaller, more distant systems such as NGC 3801 \citep{Croston2007} and a rather more complex scenario in the spiral galaxy NGC 6764 \citep{Croston2008}. We recently carried out a \textit{Chandra} observation of Markarian 6 (Mrk 6, IC 450, $z=0.018676$), an early-type S0 Seyfert 1.5 galaxy \citep{Osterbrock1976} whose characteristics have been studied in the radio ($L_{1.4GHz}=1.7\times10^{23}$ W Hz$^{-1}$ sr$^{-1}$), infrared, optical and X-ray wavelengths over the last 30 years. Recent radio studies have unveiled a complex structure surrounding the AGN, with a double set of bubbles and a radio jet \citep{Kukula1996}, suggesting a jet precession scenario \citep{Kharb2006}. In this paper we describe the results of our analysis of the \textit{Chandra} observation.

We have also analysed three previous datasets from \textit{XMM-Newton} to study the evolution of the AGN and its immediate environment over time. These data were analysed in detail previously \citep{Schurch2006, Immler2003}, using different models to address the complex scenario surrounding the AGN. We have approached this analysis by searching for consistency between the \textit{Chandra} and \textit{XMM-Newton} datasets in the context of variable absorption, following both the methods used by \citet{Hardcastle2009} when modelling a sample of 3CRR radio sources, and a double partial covering model that has previously been successful in describing the properties of the nuclear spectrum of Mrk 6.

Throughout this paper we use a concordance cosmology, with $H_0$=70 km s$^{-1}$ Mpc$^{-1}$, $\Omega_m$=0.3 and $\Omega_{\Lambda}$=0.7. The redshift of Mrk 6 corresponds to a luminosity distance of 78.4 Mpc.

%
\section{Observations and data reduction}\label{ObsAndData}

\begin{figure*}[ht!]
\centering
\includegraphics[width=0.92\textwidth]{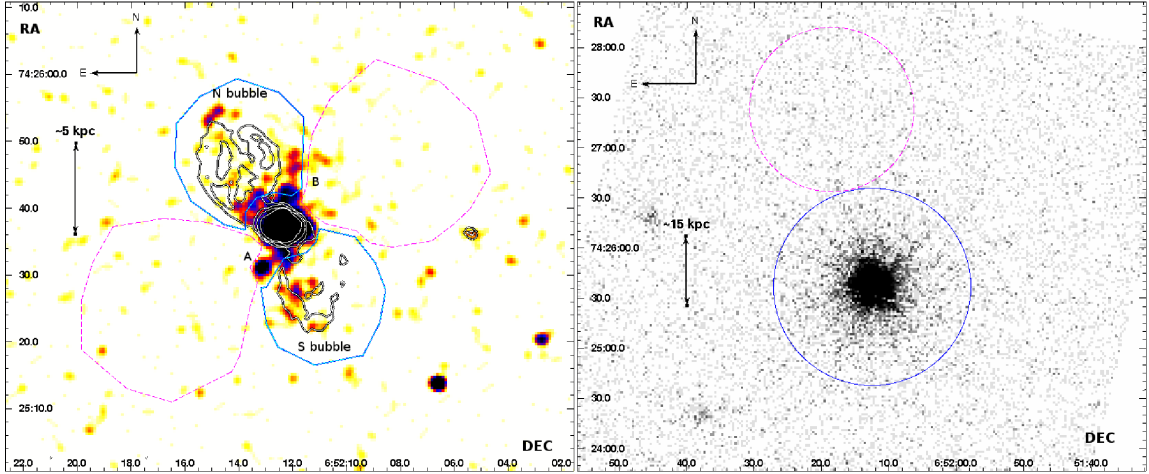}
\caption{\textbf{Left:} \textit{Chandra} ACIS-S 0.4-3 keV image of Mrk 6, with $\sigma$=3 pixels Gaussian smoothing and overlaid VLA radio contours \citep{Kharb2006}, showing the match between the radio and X-ray emission. In this frame are also displayed the source (blue, continuous) and background (magenta, dashed) extraction regions for the bubble spectra. The label A shows a possible background quasar, while B indicates the emission coincident with the extended [OIII] emission-line regions, see Figure \ref{f02}. The energy range has been selected to highlight the shell-like structure surrounding the bubbles and the overlap between the X-ray emission and the radio structure. \textbf{Right:} (non-smoothed) \textit{XMM-Newton}'s MOS 1 image (obsid 0305600501) of Mrk 6, with the source (blue, continuous) and background (magenta, dashed) extraction regions for all the AGN spectra (\textit{Chandra} and \textit{XMM}).}\label{f01}
\end{figure*}

We observed Mrk 6 with \textit{Chandra}'s ACIS-S CCD on July 12 2009, with a total exposure of 75 ksec. The observation was carried out in faint mode, to minimise issues with the background and at the same time avoid telemetry saturation. There were no intervals of high background during this observation, hence we analyse the full exposure. We reprocessed the data from level 1 events in the standard manner, using CIAO 4.2 and CALDB 4.2.0. To improve the quality of the data for the purposes of a study of the extended radio bubbles, we removed the pixel randomization and ran the destreaking routines on the data. We produced two filtered images (0.3-7 keV and 0.5-5 keV) to assess the extent and structure of the radio bubbles, and chose our regions accordingly, but extracted all the spectra from the original events file. We performed our spectral fits with XSPEC, constraining the energy range to be coincident with that covered by the instrument calibration (0.3-7 keV). 
We estimated the pileup fraction for these data from the \textit{Chandra} documentation and calculated a numerical value using the PIMMS tools. The pileup fraction is about 12\% throughout the observation (there are no substantial differences in the count rate during the observation).

The \textit{XMM-Newton} observations were taken in March 2001 (obsids 0061540101 and 0061540201), April 2003 (obsid 0144230101) and October 2005 (obsid 0305600501). We reduced the data using the standard routines from SAS version 9.0 and the latest calibration files. Table \ref{observations} gives details on the exposure times. The 2001 and 2003 observations were taken with the medium filter, while 0305600501 was taken with the thin filter. We decided not to use the observation 0061540201 because it was taken in calibration closed mode. We discarded the first 10 ksec from the observation 0144230101 due to high background. Despite the background contribution being uneven during this observation, after examining the relative count rates of the source and the background, we decided not to discard any other time intervals, since the background in these periods was always below 10\% of the intensity of the source across the whole energy range, which is accurate enough for our purposes. We extracted spectra for the PN, MOS1 and MOS2 cameras, except in the case of the observation 0144230101, where a PN spectrum could not be obtained. We limited our spectral fits to the 0.3-8.0 keV energy range, to analyse an energy range comparable to that covered by \textit{Chandra}.

For the study of the properties of the AGN we used the same extraction region for the \textit{Chandra} and \textit{XMM} data, a $\sim$60 arcsec radius circle centred in the source, which contains most of the MOS and PN PSF. Although we know from the \textit{Chandra} image that there are a few point sources other than the AGN within this extraction region (see Figure \ref{f01}), they are not resolved by the \textit{XMM} instruments, and their intensities are so low compared to the AGN that we can consider their effects negligible. We also use the same background region for all the data, a $\sim$50 arcsec radius circle, North of the source region to avoid contamination from the host galaxy, as seen on Figure \ref{f01}.

To simulate the AGN on the \textit{Chandra} detector we used the ChaRT 1.0 web interface and MARX version 4.5.

\begin{figure}[h!]
\centering
\includegraphics[width=0.46\textwidth]{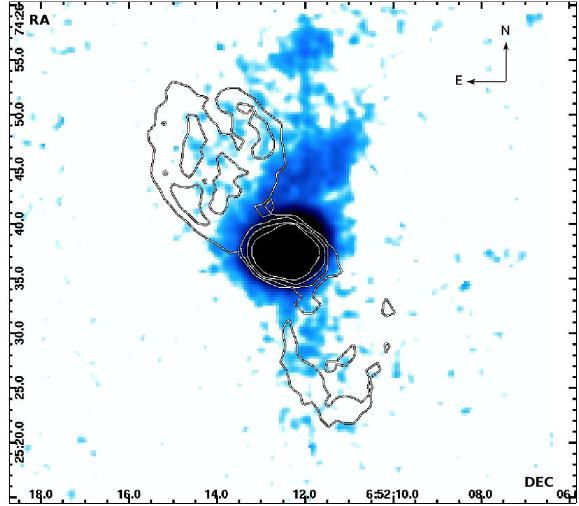}
\caption{[OIII] image of Mrk 6 from \citet{Kukula1996}, with overlaid radio contours, showing the structure of the extended emission-line region (ENLR). \textbf{Scale:} 1$\arcsec$=366 pc.}\label{f02}
\end{figure}

For the study of the extended bubble area, we used as a reference the radio data from \citet{Kharb2006}, defined contours around the largest-scale bubbles and drew source regions outside these contours, keeping well away from the central AGN to avoid contamination from its PSF, as well as from the inner radio bubbles and jet. Figure \ref{f01} shows details on the extraction regions and the structure of the extended emission from the bubbles. In this case, we took care to exclude the only likely background source (South of the source, see discussion in Section \ref{Bubbles}) in order to avoid contamination in our spectra.

All the spectra were grouped to 20 counts per bin after background subtraction, before spectral fitting, to be consistent with $\chi^{2}$ statistics. We used a fixed Galactic absorption of $N_H=6.39 \times 10^{20}$atoms cm$^{-2}$ \citep{DL1990} for all our spectral fits, and a redshift $z=0.018676$ (\textit{SIMBAD}).

\begin{table}[h]
\centering
\setlength{\extrarowheight}{2pt}
\begin{scriptsize}
\caption{Summary of Mrk 6 X-ray observations.}\label{observations}
\begin{tabular}{ccccc}\hline\hline
	Telescope &Instruments &Date &Obsid &ksec\\\hline
\textit{XMM-Newton}&PN, MOS1,2&2001/04/26&0061540101&31\\
\textit{XMM-Newton}&MOS1,2&2003/04/26&0144230101&41\\
\textit{XMM-Newton}&PN, MOS1,2&2005/10/27&0305600501&20\\
\textit{Chandra}&ACIS&2009/06/12&10324&75\\\hline
\end{tabular}
\end{scriptsize}
\end{table}

All the parameters estimated with XSPEC are quoted with 90\% confidence uncertainties. The errors on the upper limit of counts for the extended regions are $1\sigma$.

%
\section{Results}\label{results}

%

\subsection{The Seyfert core}\label{AGN}

\begin{figure*}[t!]
\centering
\includegraphics[width=0.92\textwidth]{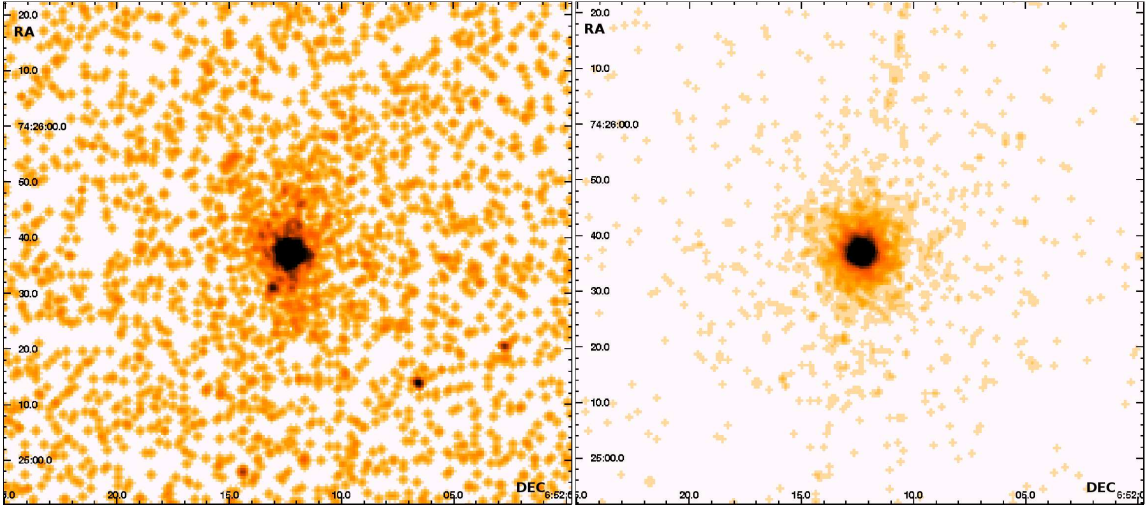}
\caption{\textbf{Left:} real \textit{Chandra} ACIS-S (75 ks) 0.3-7 keV image of Mrk 6, with $\sigma$=2 pixels Gaussian smoothing. \textbf{Right}: ChaRT/MARX simulated data (150 ks) 0.3-7 keV, with $\sigma$=2 pixels Gaussian smoothing. The scales and colour bars in both pictures are the same. The simulated image contains only AGN photons, with no background, and helps us correct the pileup effects in our AGN spectra (we did not simulate pileup) and establish a statistical contribution of photons from the AGN to the bubbles and external regions of the galaxy. Since the exposure time of the simulation is twice as long as the one of the original data, we have corrected for any exposure-dependent effects when using it for our analysis. The equivalent of the faint streak in the simulated image is below the background level in the real data. \textbf{Scale:} 1$\arcsec$=366 pc.}\label{f03}
\end{figure*}

\subsubsection{Instrumental corrections}\label{pileup}

When we first approached the analysis of our data, we extracted a spectrum using the regions illustrated in the right panel of Figure \ref{f01} and fitted a model consisting of a double power law with Galactic and heavy intrinsic absorption. We were surprised to find that the \textit{Chandra} and \textit{XMM-Newton} spectra looked radically different. After checking the observation details of the \textit{Chandra} data, using the on-line tool \textit{PIMMS} , we found that the AGN spectrum had a noticeable pileup fraction ($\sim$12\%), compatible with the frame time of the observation (3.1 s) and the background-subtracted count rate of the AGN ($\sim$0.11 counts/s). This resulted in a ``ghost" peak around 2 keV in our spectra, caused by piled up 1 keV photons. We could not obtain satisfactory results from the XSPEC \textit{pileup} model, perhaps due to the complexity of our underlying model, so we decided to attempt a different approach, previously applied with success by \citet{Getman2005}, \citet{Evans2005}, and \citet{Hardcastle2006}.

In most sources only the central pixels of the PSF are affected by pileup. These innermost pixels also contain a large fraction of the total photon counts; so eliminating them effectively removes any issues with pileup, but it also results in a lower signal to noise ratio. If pileup is significant enough to cause a bias in the spectral fits, however, removing its effects often results in improved fits. In our case, excluding the central four pixels of the PSF effectively eradicates any pileup issues in our spectra; we extracted a spectrum from such a region, centered on the core, and generated an RMF and ARF in the standard way. However, since the PSF is energy-dependent, excluding these pixels and extracting a spectrum from the resulting annular region alone does not solve the problem; it is necessary to correct the ARF to account for this energy dependence.

To correct the ARF we followed the procedure described by \citet{Hardcastle2006}. We generated an energy (keV) versus flux (photons/cm$^2$/s) table from our model fit and fed it to ChaRT, the Chandra Ray Tracer \citep{ChaRT}, a code that simulates a PSF from a specified model. We selected an exposure time of 150 ks, twice as long as our real exposure, to get the best possible photon statistics in the simulated data while trying not to exceed the limit on the ray density. We did not simulate the pileup with ChaRT. The output from ChaRT is a set of rays that cannot be directly used for analysis: they need to be projected on the detector plane and converted into an events file using MARX\footnote{See http://space.mit.edu/CXC/MARX/docs.html}.

We then generated a new events file from our original data and an annular extraction region, identical to the one we used to generate our spectra, but excluding the central four pixels. We used a code to fit a 5-th degree polynomial to the ratio of this events file and the whole simulated events file as a function of energy. This code reads in the ARF generated by CIAO and scales the effective area at each energy, using the polynomial fit, to effectively correct for the missing effective area due to the exclusion of the central pixels. The code then writes a new ARF which can be used to correct for the effects of excluding the central pixels.

We were then able to carry out fits to our extracted spectrum. As expected, the loss of counts from the central pixels slightly decreased the signal to noise ratio, but eliminating the bias caused by pileup resulted in a noticeably improved fit (see Section \ref{Models} for details) and we obtained a better statistical result. The simulated image is displayed on the right panel of Figure \ref{f03}, next to the original data. Although the simulation predicts some streaking in the image, in our real exposure it is so weak it is not detectable over the background noise, so that we cannot use it to constrain the AGN spectrum.

\subsubsection{Models}\label{Models}

Early \textit{ASCA} data on Mrk 6 have already made clear the difficulties underlying the study of intermediate type Seyferts \citep{Feldmeier1999}, particularly when disentangling the absorption component from the intrinsic continuum shape, and several attempts have been made since then to successfully model both the underlying, complex physical scenario, and the properties of the X-ray emission we see. The partial covering model used for the \textit{ASCA} data was also employed in the analysis of an early \textit{XMM-Newton} dataset by \citet{Immler2003}, while analysis of a later observation \citep{Schurch2006} favoured the inclusion of a reflection component. 

After correcting for pileup in our \textit{Chandra} spectrum, it still looked very different to the \textit{XMM-Newton} spectra. Moreover, we encountered some problems when fitting the \textit{Chandra} spectrum alone, using a double power law model with local and intrinsic absorption (see below for a detailed description of the model). XSPEC is unable to disentangle the contribution to the model from the power law from that of the absorbing column, which results in extremely low or even negative values of the photon index for the second power law. Seeing that the \textit{Chandra} spectrum has fewer relative counts and a different shape in the 2-5 keV range, we decided to test the hypothesis of a variable absorption column along our line of sight to Mrk 6, which may be caused by the movement of clumpy gas from the regions within a few pc of the black hole, happening on timescales of months to years. This hypothesis has been successfully used over a variety of Seyfert 1 and 2 X-ray spectra (see e.g. \citeauthor{Risaliti2002} \citeyear{Risaliti2002}) and has already been suggested before for Mrk 6 by \citet{Immler2003} after they observed a substantial change in the absorption column between their \textit{XMM} and \textit{BeppoSAX} observations. However, the absorption variations between the \textit{XMM} spectra from 2001, 2003 and 2005 are much smaller \citep{Schurch2006}. Disentangling intrinsic AGN variability from the variable obscuration is a difficult task, both in the X-rays and at other wavelengths. We know from the optical wavelengths that changes in intensity of the H$\alpha$ and H$\beta$ optical emission lines are a good probe of the variability of the central AGN itself, but there is also evidence that suggests that the gas where these lines are produced can undergo substantial variations over time \citep{Rosenblatt1992}. 

\begin{figure}[h!]
\centering
\includegraphics[width=0.46\textwidth]{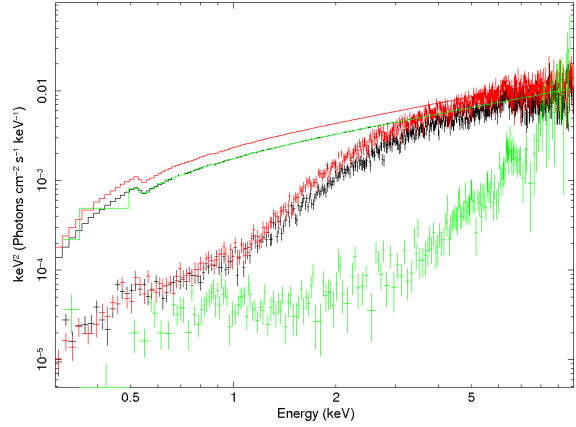}
\caption{Energy-unfolded \textit{XMM-Newton} PN (black: 0061540101; red: 0305600501) and \textit{Chandra} ACIS-S (green) spectra of Mrk 6 to illustrate the changing spectral shape caused by the variation in N$_H$. The model is a power law with local absorption, with the photon index and normalization fixed to the values obtained in the fit of the second model (see Table \ref{joint}).}\label{fig04}
\end{figure}

We decided to approach the modeling of the AGN both from the perspective used in \citet{Hardcastle2009} and \citet{Croston2007}, which has proved successful in describing the properties of many powerful radio sources at low to intermediate redshifts, and under the partial covering models that have been successful at describing not only the properties of the spectrum of Mrk 6 \citep{Feldmeier1999,Immler2003,Schurch2006}, but also several other classical Seyferts. 

\begin{figure}[h!]
\centering
\includegraphics[width=0.46\textwidth]{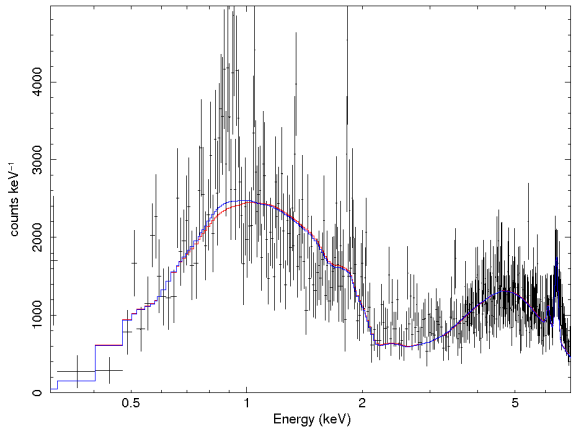}
\caption{\textit{Chandra} AGN spectrum with the best fitting parameters for Model 1 (see Table \ref{joint} for details). The model in red does not take into account the thermal emission from the radio bubbles, the model in blue does, and fits the data better around 1 keV. The addition of the thermal component increases the model predicted flux by 12\% in the 0.7-1.1 keV range.}\label{f05}
\end{figure}

The first model provides an accurate description of the nuclear spectrum of Cen A \citep{Evans2004} and is often used to fit spectra of narrow-lined radio galaxies. We fitted a single power law with a fixed Galactic absorption to the data ({\it wabs} XSPEC model). We added to this model a second absorption column ({\it zwabs} XSPEC model) at the source's redshift, which was allowed to vary, and a second power law. We added a redshifted Gaussian to account for the Fe K$\alpha$ emission line. See Figure \ref{f05} for an overview of the \textit{Chandra} spectrum.

\begin{landscape}

\begin{table}[ht]
\begin{scriptsize}
\begin{center}
\setlength{\tabcolsep}{1.5pt}
\setlength{\extrarowheight}{5pt}
\caption{Best fitting parameters for the \textit{Chandra} and \textit{XMM-Newton} AGN fit. \textbf{Model 1:} wabs(apec+po+zwabs(po+zgauss)); \textbf{Model 2:} wabs(apec+zpcfabs$_1$*zpcfabs$_2$(po+zgauss)). Both models include the contribution from the thermal emission from the radio bubbles (see Section \ref{Bubbles} for details).}\label{joint}
\begin{tabular}{c|cc|c|cc|ccc|ccccccc}\hline\hline
&&&&\multicolumn{2}{c}{apec}&\multicolumn{3}{c}{Model 1}&\multicolumn{7}{c}{Model 2}\\
	Obsid& Line E&Eq. Width&Local $N_H$&kT&abundance&$\Gamma_1$&$N_H$&$\Gamma_2$&f$_{cov}^1$&$N_H^1$&f$_{cov}^2$&$N_H^2$&$\Gamma$&Norm&L$_{2-10 keV}$\\
	&keV&eV&$\times 10^{20}$at. cm$^{-2}$&keV&$\odot$&&$\times 10^{22}$at. cm$^{-2}$&&&$\times 10^{22}$at. cm$^{-2}$&&$\times 10^{22}$at. cm$^{-2}$&&$\times 10^{-3}$ ph. keV$^{-1}$ cm$^{-2}$ s$^{-1}$&$\times 10^{43}$erg s$^{-1}$\\\hline
10324&6.46$^{+0.06}_{-0.07}$&$309^{+331}_{-207}$&6.39&$0.87^{+0.25}_{-0.23}$&0.35&1.5&30.78$^{+3.28}_{-4.57}$&1.16$^{+0.03}_{-0.03}$&0.68$^{+0.01}_{-0.01}$&17.93$^{+3.04}_{-2.54}$&0.83$^{+0.01}_{-0.01}$&55.89$^{+5.46}_{-4.70}$&1.28$^{+0.01}_{-0.01}$&1.94$^{+0.12}_{-0.08}$&1.23\\
0061540101&6.43$^{+0.02}_{-0.03}$&$94^{+36}_{-32}$&6.39&$0.87^{+0.25}_{-0.23}$&0.35&1.5&2.70$^{+0.08}_{-0.08}$&1.16$^{+0.03}_{-0.03}$&0.68$^{+0.01}_{-0.01}$&0.87$^{+0.05}_{-0.05}$&0.83$^{+0.01}_{-0.01}$&3.32$^{+0.09}_{-0.08}$&1.28$^{+0.01}_{-0.01}$&1.99$^{+0.03}_{-0.02}$&1.25\\
0144230101&6.40$^{+0.05}_{-0.05}$&$62^{+60}_{-58}$&6.39&$0.87^{+0.25}_{-0.23}$&0.35&1.5&0.80$^{+0.05}_{-0.05}$&1.16$^{+0.03}_{-0.03}$&0.68$^{+0.01}_{-0.01}$&0.0&0.83$^{+0.01}_{-0.01}$&0.90$^{+0.03}_{-0.03}$&1.28$^{+0.01}_{-0.01}$&2.05$^{+0.02}_{-0.02}$&1.29\\
0305600501&6.42$^{+0.03}_{-0.04}$&$53^{+31}_{-29}$&6.39&$0.87^{+0.25}_{-0.23}$&0.35&1.5&2.38$^{+0.07}_{-0.07}$&1.16$^{+0.03}_{-0.03}$&0.68$^{+0.01}_{-0.01}$&1.05$^{+0.05}_{-0.05}$&0.83$^{+0.01}_{-0.01}$&2.38$^{+0.06}_{-0.06}$&1.28$^{+0.01}_{-0.01}$&2.64$^{+0.03}_{-0.03}$&1.66\\
\multicolumn{6}{c}{}&\multicolumn{3}{c}{$\chi^{2}$=3116 / 2980 DOF}&\multicolumn{7}{c}{$\chi^{2}$=3046 / 2978 DOF}\\\hline
\end{tabular}
\end{center}
\end{scriptsize}
\end{table}

\end{landscape}

\begin{figure}[h!]
\centering
\includegraphics[width=0.46\textwidth]{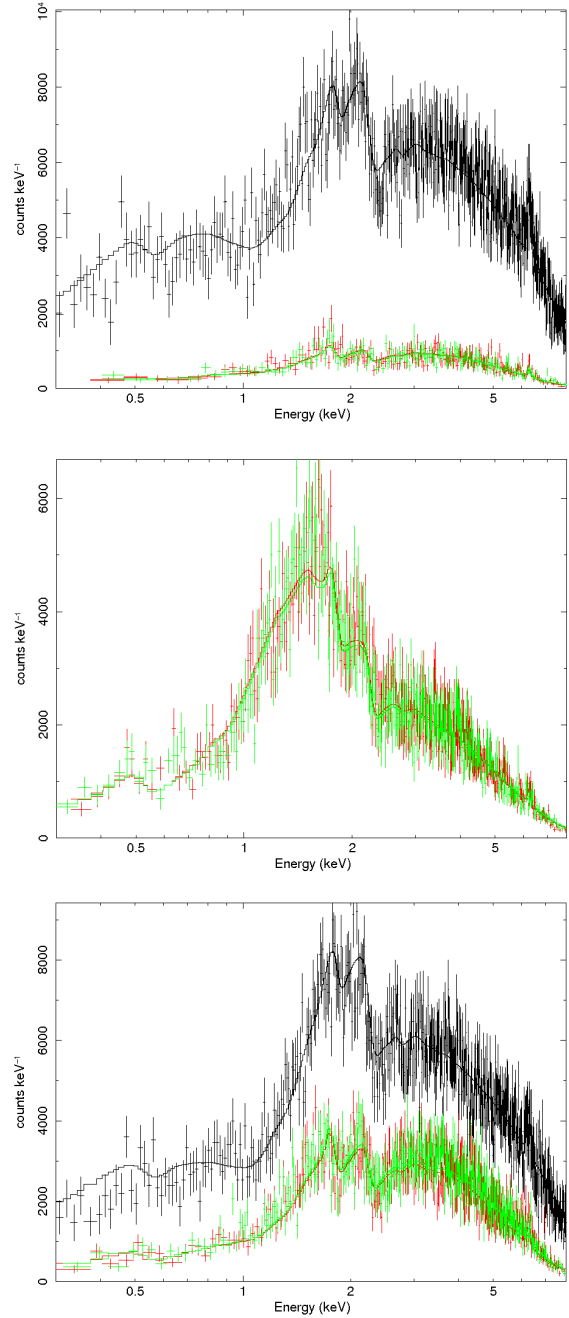}
\caption{\textit{XMM} AGN spectra with the best fitting parameters for Model 1 (two power laws + intrinsic absorption + thermal emission from the bubbles, see Table \ref{joint} for details). From top to bottom, obsids 0061540101, 0144230101, 0305600501, with the PN spectrum in black, MOS1 in red and MOS2 in green.}\label{f06}
\end{figure}

Although the \textit{Chandra} data have the spatial resolution required to differentiate the AGN contribution from the extended thermal emission of the radio bubbles, we used the same extraction regions on all our spectra (see Figure \ref{f01}), so as to be able to make a direct comparison between them. We decided to take into account the contribution of the thermal emission from the bubbles in our spectra by adding the {\it apec} XSPEC component to our model. We will discuss the details and implications of this extended emission in Section \ref{Bubbles}; an overview of its effect on our AGN spectrum can be seen in Figure \ref{f05}.

We individually fitted all the spectra, for each instrument of the \textit{XMM} and \textit{Chandra}, and then tied together the parameters for the \textit{XMM} spectra belonging to the same observation, to assess the number of variable parameters: we assumed that the normalizations do not vary within the same observation between the MOS and PN instruments, and tied together the values for the three instruments, with no significant impact on the final statistical result. We finally attempted a full joint fit with four groups of data, one for \textit{Chandra} and one for each \textit{XMM} observation. When setting the variable parameters for the joint fit, we also decided that the photon indices for the second power law should be tied across the datasets, to break possible degeneracies with the local absorption columns, and froze the first photon index to a reasonable value (1.5) to minimise the overlap degeneracy between both power laws. This allowed us to make the joint (\textit{Chandra} + \textit{XMM}) fitting considerably faster. See Figures \ref{f05} and \ref{f06} for an overview of the \textit{Chandra} and \textit{XMM} spectra and Table \ref{joint} for details on the fit parameters for this model (Model 1).

The initial fit to the data resulted in $\chi^{2}$=3331 for 3149 degrees of freedom (reduced $\chi^{2}$=1.058). After correcting for the pileup effects, with the methods described in Section \ref{pileup}, we obtained $\chi^{2}$=3125 for 2980 degrees of freedom (reduced $\chi^{2}$=1.049). After adding the contribution from the radio bubbles, with the parameters frozen to their best fit values, we obtained $\chi^{2}$=3116 for 2980 degrees of freedom (reduced $\chi^{2}$=1.046). Since the corrections mainly affect the \textit{Chandra} spectrum their contribution to the global result is quite small, but not negligible.

As shown in Table \ref{joint}, we obtained a combined value for the index of the second power law of 1.16$^{+0.03}_{-0.03}$, which is quite flat. This is partly caused by the difficulty of disentangling the index of the power law from the heavy absorption in the \textit{Chandra} spectrum, as mentioned above. We find and fit the Fe K$\alpha$ emission line, setting $\sigma$=0 to simplify the fit. The best fitting energies and equivalent widths for the emission line are shown in Table \ref{joint}. The line energy is consistent for all the datasets, although the intensity of the line does change with the AGN state and the local absorption column, as expected. The errors on the equivalent widths are quite big, which may be a sign of the line having some breadth, as suggested by \citet{Feldmeier1999}, or be a side effect of the complicated model.

To compare our results with the previous \textit{ASCA} and \textit{XMM} results, and verify our hypothesis of variable absorption, we tested a few other models. We found that a single power law, with local absorption, or local + intrinsic absorption, results in values similar to those obtained by \citet{Immler2003}, although it gives a very poor fit to the 2003 \textit{XMM} spectra and does not fit at all the \textit{Chandra} data (reduced $\chi^2$ $\sim$6). This is most likely caused by the varying absorption column, which cannot be properly accounted for with such a simple model.

We found that the only model that could allow a direct comparison with all of the previous results was one consisting of local absorption, a double partial covering ({\it zpcfabs}), a power law and a Gaussian to account for the Fe emission line. We tied together the covering fractions across the different spectra, allowing only the $N_H$ to vary. We did not add any reflection components to this model, since they are not considered in the \textit{ASCA} analysis by \citet{Feldmeier1999}, but we did add the contribution due to the thermal emission from the radio bubbles. The results of this fit are shown in Table \ref{joint} under the label ``Model 2". We find that this model gives $\chi^{2}$=3046 for 2978 degrees of freedom, reduced $\chi^{2}$=1.023. We found that the photon index of the power law in our fit is lower than the one obtained in previous analysis of Mrk 6, only similar to the value obtained by \citet{Schurch2006} with their third model. This is partly due to the addition of the thermal emission component, which we deem necessary since it has been spatially resolved and fitted to the \textit{Chandra} data and results in a statistical improvement in our other model. It is also partly caused by the effect of the variable absorption. We find the values of the Fe K$\alpha$ emission line energy and equivalent width to be consistent with the ones we obtained for the previous model. We found that it is necessary for the two partial covering components to overlap ($f_{cov}^{1}+f_{cov}^{2}>1$) for the fit to be statistically acceptable. We tried to fix the values to the ones employed by \citet{Schurch2006} in their analysis, but this results in a reduced $\chi^{2} \sim$1.7.

While \citet{Immler2003} cite a value for this photon index of 1.81 from their analysis of the \textit{Beppo-SAX} data, which is less affected by variable absorption, they mention in their analysis that the source was in a higher state when those data were obtained. Another \textit{Beppo-SAX} observation analysed by \citet{Malizia2003} gives a photon index $\sim$1.5. \citet{Schurch2006,Feldmeier1999} also favour a higher value for the photon index (1.6-2.2), but setting the photon index to 1.6 results in a reduced $\chi^{2}\sim$1.13 in our joint fit, and the fit requires a heavier absorbing column for the second partial covering ($N_H=66.61^{+4.92}_{-4.74}\times10^{22}$ cm$^{-2}$   for the \textit{Chandra} spectrum).

As shown in Table \ref{joint}, the absorption column is over an order of magnitude larger in the \textit{Chandra} data than in the \textit{XMM}, for both models. This high value is more similar to the ones obtained by \citet{Feldmeier1999} in their analysis of the \textit{ASCA} data. The hypothesis of ``obscuring" clouds near the AGN was proposed by \citet{Immler2003}, and \citet{Schurch2006} supported it after obtaining a relatively smaller absorption column in their analysis of the \textit{XMM} data from 2003. This detection by \textit{Chandra} rules out instrumental bias. Although more data are needed to make a statistical test, current evidence seems to point towards denser material having been observed in 1997 (\textit{ASCA}) and 2009 (\textit{Chandra}), with the passing clouds being least dense in 2003 and intermediate obscuration values in 2001 and 2005 \textit{XMM-Newton}. This allows us to establish a variability timescale of 2-6 years, which translates to a distance to the central black hole of just a few pc. Since grating data would be very useful to get an additional insight on the structure and composition of this variable absorption, we briefly investigated the \textit{XMM-Newton} RGS spectra from the three observations. Unfortunately, the signal to noise is rather bad, and no evident emission or absorption lines can be inferred from them, as already noted by \citet{Schurch2006} and \citet{Immler2003}.

Although the statistical result is better for the second model, we must consider the physical implications of both models. This AGN seems to fall between the behaviour expected from a classical Seyfert and that of a radio galaxy, it is therefore interesting to explore the consequences of these results in the light of what we know about these classifications. As explained in \citet{Hardcastle2009}, the soft excess characteristic of most Narrow-Line Radio Galaxies (NLRGs) can be explained by three classes of model:
\begin{itemize}
\item{It is thermal or line emission, from the photoionized material close to the AGN or from the IGM of the host galaxy.}
\item{It is non-thermal, power law emission from the central AGN, visible through the partial covering material.}
\item{It is non-thermal, power law emission related to the jet.}
\end{itemize}

The first model can most likely be discarded in this case, since neither the \textit{Chandra} nor the \textit{XMM} spectra show evidence for strong emission lines. The RGS spectra do show some residuals over a pure continuum model in the 0.5-1 keV region that could be a hint of oxygen emission lines \citet{Schurch2006}, but they are quite faint, and since the area is not spatially resolved, they could be due to the thermal emission from the radio lobes.

To test which of the two other models is most likely to apply in this case we calculated the X-ray luminosity of the unabsorbed power law from our first model, $L_{X,u}=9.6 \times 10^{41}$ erg s$^{-1}$, and the $\nu F_{\nu}$ 5GHz luminosity at the base of the jet from our radio maps, $L_{5GHz}=1.41\times10^{38}$ erg s$^{-1}$. On the plots by \citet{Hardcastle2009} this falls quite far from the behaviour expected for radio galaxies. We must therefore conclude that the jet-related soft emission must be quite low, and the dominant contribution to the soft excess is most likely the emission from the central AGN.

Following the results of \citet{Evans2010} we also applied their model to our data, to check if it would fit a photon index for the power law more consistent with the values expected from other intermediate Seyferts. This model tests a different absorption component for the soft and hard power laws: {\it wabs(apec+zwabs*po+zwabs*zpcfabs(zgauss+po))}. We obtained a fit to the data similar to that of our previous partial covering model, $\chi^{2}$=3037.22 for 2971 DOF, reduced $\chi^{2}$=1.022, and a similar photon index for the second power law, 1.25.

We also tested a model with a multiphase medium. We found that a single warm absorber does not provide a good fit to the data, and hence split the intrinsic partial covering absorptions of our second model into a cold ({\it zpcfabs}) and a warm ({\it zxipcf}) component. This model seems to provide a better fit for the soft end of the spectra, and yields a $\chi^{2}$=3000.98 for 2977 DOF, reduced $\chi^{2}$=1.007. The absorbing columns are similar to those we found for the double cold partial covering, and the photon index of the power law is 1.41$^{+0.04}_{-0.03}$, higher than with any other model we have tested. However, the ionization parameter is not very well constrained, $log \Xi=-0.09^{+0.38}_{-0.47}$, since the model is limited by the resolution of our spectra. Warm absorbers have been observed in other Seyferts, albeit at higher ionization values (see e.g. \citeauthor{Longinotti2009} \citeyear{Longinotti2009}), but in our case the situation is less clear. The soft excess this model successfully accounts for can have several origins: the jet, warm-hot gas from other regions of the galaxy, or a higher normalization of the thermal emission from the lobes, since we decided to freeze the latter to the maximum value provided by the spatially resolved fit. Unfortunately the grating spectra are not sensitive enough to test either hypothesis, but future, longer exposures may hold the key to the complexity we are observing in Mrk 6.

%

\subsection{Radio bubble-related emission}\label{Bubbles}
To extract the spectra of the extended radio bubbles we only used the \textit{Chandra} data, since in the \textit{XMM} images this area is completely dominated by the PSF of the central AGN. We used the contours from the radio images obtained by \citet{Kharb2006} to define our extraction regions around the 7-kpc radio bubbles, which can be seen in Figure \ref{f01}. The photon statistics for the bubbles are not very good; we do not have enough counts to fit to both bubbles separately and determine their temperatures independently, and even for the joint extraction the photon count is quite low ($\sim$200 counts after background subtraction, see Figure \ref{f07}). Even though we were careful to keep these regions well away from the central AGN, there is still some contamination from its PSF in our spectra. The data we simulated with ChaRT and MARX to correct the effects of pileup in the AGN spectrum (see Section \ref{AGN}) show the number and energy distribution of photons from the AGN PSF across all the CCD, see Figure \ref{f03}, and can thus be used to estimate the AGN photon contribution to our spectra from these external regions. Our calculations show that this contamination amounts to up to 60\% of the counts in the spectra prior to background subtraction. However, the background regions that we have chosen contain a similar number and energy distribution of photons from the AGN, its PSF being quite homogeneous on these scales, and their effect is hence mostly eliminated. We carried out a study of the AGN contamination in our regions by studying the energy distribution of the AGN counts in our simulated data, within the same extraction regions. We found most of the counts to be contributing at energies above 4 keV. There is a minor contribution at energies slightly above 1 keV. The residuals we get from our fit at these energy ranges are likely to have an origin in this contamination (see Figure \ref{f07}).

We have found a source of contamination from high energy photons in the form of a small bright region ($L_{X}\sim1.5\times10^{39}$ erg s$^{-1}$) that can be seen on the left panels of both Figure \ref{f01} and Figure \ref{f03} just South of the source (labeled as ``A" in the left panel of figure \ref{f01}). We plotted a histogram of the counts from this region, and by their distribution, and the fact that it has no radio or optical counterpart, we infer this is clearly a point source, most likely a background, more distant quasar. It is quite unlikely that this object may be an X-ray binary (XRB), but it could be an ultra-luminous X/ray source (ULX), since its magnitude should be above our detectability threshold for this distance ($L_{ULX}\sim 10^{39}-10^{41}$ erg s$^{-1}$) although the morphology of the host galaxy indicates this is not very likely \citep{Swartz2004}. 

The other bright structure North of Mrk 6, labeled as ``B" in the left panel of Figure \ref{f01}, does seem to have a thermal origin, and a similar energy histogram as the photons we find in the bubbles; part of it is clearly surrounding the radio structure, and we therefore assume that it is part of the hot shell outside the radio bubbles and include it in our extraction regions. It must be noted, however, that Mrk 6 has an inner set of $\sim$1.5 kpc radio bubbles, almost perpendicular to the outer $\sim$7.5 kpc structure (see \citeauthor{Kharb2006} \citeyear{Kharb2006}), and this is likely to cause some turmoil in the gas surrounding the AGN. These inner bubbles are not resolved by  \textit{Chandra}. The jet of Mrk 6 was detected by \textit{MERLIN} \citep{Kukula1996} and is aligned in the North-South direction, so that this is the presumed photoionizing axis of the AGN.

This direction also corresponds to the Extended Narrow-Line Region (ENLR) of Mrk 6, see Figure \ref{f02} and \citet{Capetti1995,Kukula1996,Kharb2006} for details. An overlap between [OIII] and X-ray emission has been observed in many Seyfert 2 AGNs before, and it is believed to happen when the X-ray photons heat up and ionize the cold gas around the AGN. The [OIII] emission in this system seems to correspond to different structures: the emission closer to the active nucleus is clearly caused by direct photoionization from the AGN, while the two bulges that extend in the N-S direction, showing both enhanced [OIII] and X-ray emission, are most likely caused by the interaction of the current jet with the surrounding material, which heats it up both through shocks and photoionization (see \citeauthor{Capetti1995} \citeyear{Capetti1995} and \citeauthor{Kukula1996} \citeyear{Kukula1996} for details on these inner structures). Since the overlap between the [OIII] and X-ray emission can also be observed at kpc distances, extending to the North of the AGN, we calculated whether the nucleus could be photoionizing these regions to X-ray and [OIII] emitting temperatures, following the steps of \citet{Wang2009} and the models from \citet{Kallman1982}. For the X-ray emission that can be seen $\sim$3 kpc North of the AGN, around the edge of the radio structure, we used the parameters derived from our fits of the nuclear spectrum and the radio lobes. We found that such X-ray emission cannot be caused by photoionization from the AGN, since the unabsorbed luminosity of Mrk 6 ($\sim1.3\times10^{43}$ erg s$^{-1}$) can only account for photoionized X-ray emission at distances within 300 pc of the central source. The [OIII], however, could be photoionized by the AGN at much larger distances, as long as the electron density stays low ($<10^{-1}$ cm$^{-3}$ for distances over 3 kpc). It is likely that some of the [OIII] emission to the North of the source may have been caused by the shock itself, both via direct heating and photoionization. The fact that the [OIII] emission follows the outline of the radio structure, but is mostly detached from it, also supports this scenario.

{With our photon statistics we cannot rule out a model in which this emission is produced in some way by a precessing jet, which would also explain some of the emission in the ENLR. Both assumptions could explain the structures that surround the real, but not the simulated AGN, on Figure \ref{f03}. For details on the radio structure of Mrk 6 and the discussion of a possible jet precession scenario see \citet{Kharb2006}.

\begin{figure}[h]
\centering
\includegraphics[width=0.46\textwidth]{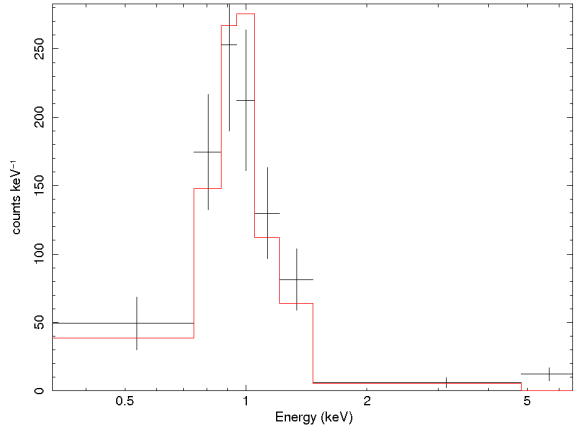}
\caption{Spectral fit to the spatially constrained radio bubble regions, with the best fitting \textit{apec} (T=$0.94^{+0.12}_{-0.19}$ keV, abundance=0.35$\odot$, $\chi^{2}$=9.39 for 8 PHA bins and 6 DOF).}\label{f07}
\end{figure}

We fitted to our data an emission spectrum from collisionally-ionized diffuse gas ({\it apec} model in XSPEC) with the abundance fixed at 0.35$\odot$ and Galactic absorption, using again the averaged column density result $N_H=6.39 \times 10^{20}$atoms cm$^{-2}$ from \citet{DL1990}. We estimated a temperature for the bubbles of $0.94^{+0.12}_{-0.19}$ keV. This model yields a $\chi^{2}$ value of 9.39 for 8 PHA bins and 6 degrees of freedom. Different extraction regions yield slightly lower values of $kT$, but still well within the errors, and different values of $\chi^{2}$, due to the reduced number of bins. We also tested a larger abundance of 0.6$\odot$ consistent with our estimations on the thermal emission of Cen A \citep{Kraft2003}, but this requires a lower normalization and yields a poorer fit to the data ($\chi^{2}$ of 11.42 for 8 PHA bins and 6 DOF).

The normalization of this component is $8.5^{+1.8}_{-1.8}\times10^{-6}$cm$^{-5}$, just high enough to be noticeable on the spectra we extracted for the AGN ($\sim$2\% of the flux between 0.8-1.1 keV). We added an \textit{apec} component to our AGN model and performed a statistical test on the temperature, finding it to agree with the results from the spatially constrained spectrum ($kT=0.87^{+0.25}_{-0.23}$ keV). This addition also improves the statistics of the joint fit, if only slightly (reduced $\chi^{2}$ reduced from 1.051 to 1.049). The effects of adding this component to the \textit{Chandra} spectrum can be seen on Figure \ref{f05}.

To discard the possibility of the emission being non-thermal synchrotron or inverse Compton, we attempted to fit a model consisting of a power law and local absorption to the data, and obtained very poor results (reduced $\chi^{2}$ of 3.79). Therefore we can quite safely assume the emission to be thermal.

Although the abundance that best describes the thermal emission from the bubbles is 0.35$\odot$, more consistent with NGC 3801 (0.3-0.4$\odot$) than with Cen A (0.6$\odot$), the statistical difference alone is not enough to rule out a higher value for the abundance, due to the low number of photon counts. We could argue that we are studying a quite small early-type S0 galaxy which probably has, as our results show, a very low gas density outside the central regions. The dust lane that is hinted at in the optical images may be a result of a past merger, there is an equivalent structure in NGC 3801. Furthermore, we must consider the influence of the active nucleus on the star formation history of Mrk 6. While it has been argued for more powerful systems that the AGN activity may indeed be suppressing star formation in the host galaxy (see e.g. \citeauthor{McNamara2007} \citeyear{McNamara2007}, \citeauthor{Schawinski2009} \citeyear{Schawinski2009}, \citeauthor{Nesvadba2010} \citeyear{Nesvadba2010}) the situation may be more complex for lower power systems, and early type galaxies in particular, as suggested by \citet{Schawinski2010}. While these arguments might be enough to support the low abundance we are observing, our tightest constraint comes from imposing the shock conditions for the bubbles. The details of these constraints and their consequences will be discussed in Section \ref{Physics}.

Even though the data show a significant correspondence between the radio contours and the X-ray emission inside the North bubble (see Figure \ref{f01}), which could be associated with jet/ISM interactions leading to the heating of gas in these regions, as in NGC 6764 (\citeauthor{Croston2008} \citeyear{Croston2008b}, \citeauthor{Kharb2010} \citeyear{Kharb2010}), most of the X-ray photons seem to correspond to the edges of the bubbles or even the regions just outside them, which is more compatible with the presence of shells of strongly shocked gas outside the bubbles, just as in NGC 3801 \citep{Croston2007}. While some structure can be seen both in the radio maps and in the X-ray image, we will assume in a first approach that the shells are spherical and uniformly filled with gas, and will later discuss the implications of a non uniform density.

The fact that the radio emission is apparently fainter in the South bubble (see Figure \ref{f01}), and the small difference in the apparent sizes of the bubbles, the North one being slightly bigger (see the discussion of the size of the shocked shells in Section \ref{Physics}), is probably due to a different past history for both bubbles, due to the gas expanding in each direction having encountered a surrounding medium with a different density, or to an asymmetric energy output from the AGN. The radio images show there is also a similar asymmetry in the inner E-W bubbles \citep{Kharb2006} and the N-S current jet \citep{Kukula1996}. The [OIII] emission is also asymmetric (see Figure \ref{f02}). These asymmetries are not visible in the optical (\textit{HST}) images of the host galaxy, the apparent dust lane being roughly perpendicular to the outer radio bubbles, but if the jet is indeed precessing, it could have created a somewhat cluttered environment around the AGN. A precessing jet is not likely to explain an asymmetry in the energy fed by the AGN to each bubble, but we cannot rule this out. Doppler boosting has been suggested for some Seyfert jets such as the one in Mrk 231 \citep{Reynolds2009}, but it is unlikely to be the cause of the asymmetry of the lobes in Mrk 6, since it would require relativistic speeds for the gas in the shells, which is inconsistent with our results (see Section \ref{Physics}). This asymmetry only underlines how challenging it is to make a self-consistent model for this kind of object. We will assume that these effects are minor for our purposes. 

%

\subsection{Luminosity constraints on the external environment}\label{Outer}

\begin{figure*}[ht!]
\centering
\includegraphics[width=0.9\textwidth]{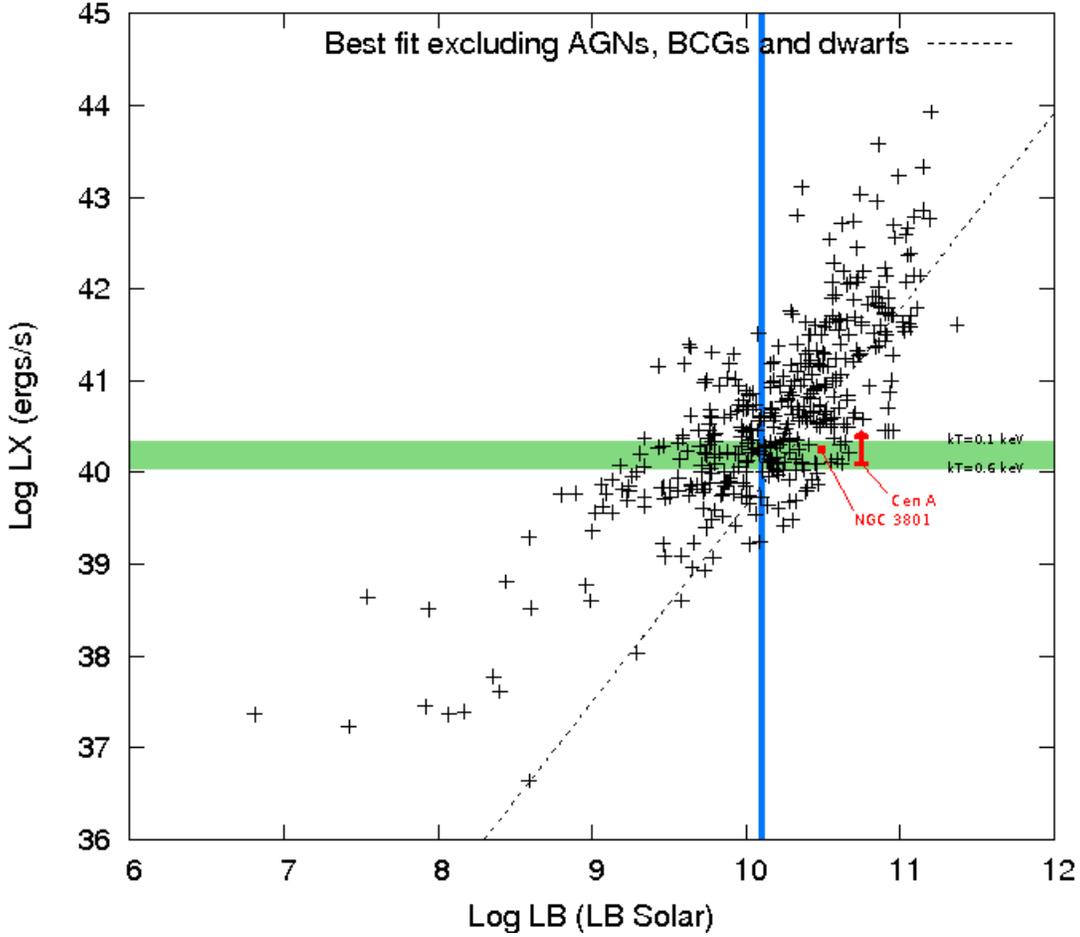}
\caption{$L_{\rm X}$ vs.\ $L_{\rm B}$ plot from \citet{Osullivan2001} with superimposed Mrk 6 B band bolometric luminosity (vertical blue line) and the expected (upper limit) X-ray luminosity range we calculated for the extended thermal emission outside the shells (horizontal green strip). The diagonal dashed line is the best fit to the data excluding AGNs, BCGs and dwarf galaxies. The AGN is most likely causing a blue excess in the optical, so the B-band luminosity is an upper limit. Given that the temperature of the external medium is most consistent with 0.1-0.2 keV, an X-ray excess caused by this emission is likely to be happening.}\label{f08}
\end{figure*}

To establish whether the gas from the bubbles is driving a shock into its surroundings we need to estimate the temperature of the external medium, derive the relative pressures and impose the Rankine-Hugoniot conditions for pressure balance \citep{Landau6}. However, we have found that due to the galaxy's morphology and size, the halo regions where the gas is expanding are not bright enough in the X-rays to allow us to obtain a spectrum. While the \textit{Chandra} spectrum appears to admit an additional thermal component with $kT\sim$0.35 keV, there is no statistical improvement on the \textit{XMM} spectra from adding it; we therefore cannot state that the component is real.

To obtain an upper limit on the luminosity and normalization of the gas component in these halo regions, we calculated the background-subtracted counts from a wide annular shell centred in the AGN ($R_{int}$=21 arcsec, $R_{ext}$=59 arcsec, excluding a few point sources resolved by \textit{Chandra}). We calculated the relative contribution of counts from the AGN using the simulated dataset, applying a scaling factor obtained from the ratio of counts within the innermost regions of the PSF (excluding the pixels we used for pileup correction) in the real and simulated data, which takes into account possible energy-dependent variations. We then subtracted the AGN contribution from the total number of counts in the halo, and added the 3$\sigma$ error, obtaining an upper limit of 270 counts, equivalent to a rate of 3.65$\times10^{-3}$ counts s$^{-1}$.

Using these statistical constraints on the number of photons from the halo, we can derive upper limits on the luminosity of this region. These upper limits also constrain the possible temperature, electron density and pressure for the gas, allowing us to test the shock conditions. To achieve this, we chose a range of possible temperatures for the medium outside the radio bubbles (0.1-0.6 keV) and used XSPEC with a toy \textit{apec} model to determine the limit on the normalization corresponding to our count rate for each of our chosen temperatures. We maintained the $N_H$ and redshift parameters fixed to the appropriate values for the system (see previous sections), and the abundance fixed at 0.35$\odot$; we will discuss the implications of larger abundances in the next Section. We then derived from the flux in these models the range of possible X-ray bolometric luminosities (see Table \ref{ne_out}).

We then estimated the B band luminosity of Mrk 6 from the data in the catalogue by \citet{DeVaucouleurs1995}, obtaining a value of 1.15$\times 10^{10}L_{B}\odot$. Since Mrk 6 contains an AGN, it is likely that we are observing a blue excess, therefore this value is to be considered an upper limit. We used as a reference the work of \citet{Osullivan2001} to estimate the feasibility of our results on the X-ray luminosity. Their work catalogues the X-ray extended emission detected by \textit{ROSAT} on several hundred early-type galaxies, and attempts to establish a relationship between the X-ray and B band bolometric luminosities for these systems. We overplotted our results on their diagram\footnote{We obtained the data from http://www.sr.bham.ac.uk/$\sim$ejos/catalogue.html} (see Figure \ref{f08}) to see whether Mrk 6 falls near or above their correlation. They excluded AGNs on their fit, and are thus only measuring the extended emission from the galaxy, which is what we are looking for in the outer regions of Mrk 6. Our results cover a wide range of possible X-ray luminosities, but especially for the lower temperatures, and considering that our estimate for the B luminosity is an upper limit, the data do fall in the right region of the plot. We can therefore assume that, although we are not directly resolving it in our \textit{Chandra} image, our upper limit of counts is consistent with the luminosity expected for the galaxy, excluding the AGN. This means these photons indeed come from the extended X-ray emission of the galaxy, not from the background or the AGN.

We also tested the K-band luminosity of Mrk 6 (1.17$\times 10^{11} L_{\odot}$) against its X-ray luminosity using the relations proposed by \citet{Mulchaey2010}, to avoid the effects of blue excess from the AGN. We found the results to agree with our estimations from the B band. Our results also agree with \citet{Mulchaey2010} on the hot gas content of a sample of early-type galaxies in different environments. Their work shows that although some of these objects seem to retain their hot gas halos when living in clusters, others lose it, while for isolated galaxies the hot gas content is mostly related to the mass of the galaxy. Mrk 6 is quite small and seems to be living in a poor environment; it is therefore likely to have lost most of its halo gas.

Finally, it must be noted that in the ChaRT/MARX simulations the wings of the PSF are known to be underestimated at large distances from the point source, especially at low energies. We contacted the \textit{Chandra} Calibration team via the HelpDesk\footnote{http://cxc.harvard.edu/helpdesk/} and consulted the available bibliography\footnote{See http://cxc.harvard.edu/cal/Hrma/hrma/psf/index.html}, and estimated that, for the regions we are considering in the halo of Mrk 6, the simulated PSF wings could be underestimated by up to a factor of 2. If this were the case, our upper limit of counts would be more conservative ($\sim$30\% lower) and we would be near the limit of non detection, implying an external gas temperature below 0.1 keV (see discussion in the next Section). If the gas were much colder than 0.1 keV we would not be able to constrain any limits on the density or assume a shock scenario. This would not be very consistent with what we know about the virial temperatures of the halo gas in early type galaxies (see e.g. \citeauthor{Naab2007}, \citeyear{Naab2007}), but since we do not have the instruments to measure the extreme UV emission a gas of these characteristics would produce, it cannot be ruled out. We must also note that the X-ray emission cannot be accounted for by photoionization mechanisms (see discussion on the [OIII] emission in the previous Section). We conclude that the shock scenario provides the most plausible explanation for the emission we are observing.

%

\subsection{Physical properties of the gas shells and ISM}\label{Physics}

\begin{table}[h]
\begin{scriptsize}
\begin{center}
\caption{Estimated \textit{apec} normalization, upper luminosity limits, electron densities and gas pressure outside the bubbles for a sample of possible gas temperatures.}\label{ne_out}
\setlength{\extrarowheight}{2pt}
\begin{tabular}{ccccc}\hline\hline
	$kT$&\textit{apec} norm&$L_X$&$n_{e,out}$&$P_{out}$\\
keV&$\times10^{-5}$cm$^{-5}$&$\times 10^{40}$erg s$^{-1}$&$\times 10^{-4} cm^{-3}$&$\times10^{-13}$Pa\\\hline
0.1&49.00&3.07&40.60&1.20\\
0.2&6.10&1.87&14.30&0.85\\
0.3&3.50&1.27&10.90&0.96\\
0.4&2.50&1.16&9.18&1.09\\
0.5&2.00&1.12&8.21&1.22\\
0.6&1.75&1.11&7.68&1.36\\\hline
\end{tabular}
\end{center}
\end{scriptsize}
\end{table}

\begin{landscape}

\begin{table*}[ht]
\begin{scriptsize}
\begin{center}
\setlength{\tabcolsep}{1.5pt}
\setlength{\extrarowheight}{5pt}
\caption{\textbf{Top:} Model normalization, estimated $n_e$ upper and lower limits for the shells, derived limits for $n_{e,out}$ required to comply with the Rankine-Hugoniot conditions for a strong shock, and inferred limits on the mass, pressure, total thermal energy, work available from the gas filling the shells and total kinetic energy. \textbf{Bottom:} Same estimations for a set of Cen A-like 200 pc-wide shells \citep{Croston2009}. The errors on the pressures and energies are derived from the errors in the temperatures of the shells in the XSPEC fit ($kT=0.94^{+0.12}_{-0.19}$ keV)}\label{ne_shells}
\begin{tabular}{cccccccccccccc}\hline\hline
	Model norm&$n_{e,N shell}$&$n_{e, S shell}$&N/S Limit to $n_{e,out}$&$M_{N shell}$&$M_{S shell}$&$P_{N shell}$&$P_{S shell}$&$E_{N shell}$&$E_{S shell}$&$W_{N shell}$&$W_{S shell}$&$K_{N shell}$&$K_{S shell}$\\
	$\times10^{-6}$cm$^{-5}$&$\times10^{-2}$cm$^{-3}$&$\times10^{-2}$cm$^{-3}$&$\times10^{-3}$cm$^{-3}$&$\times10^{7}M\odot$&$\times10^{7}M\odot$&$\times10^{-12}$Pa&$\times10^{-12}$Pa&$\times10^{56}$erg&$\times10^{56}$erg&$\times10^{55}$erg&$\times10^{55}$erg&$\times10^{55}$erg&$\times10^{55}$erg\\\hline
6.71&1.44&1.76&3.59/4.40&2.79&2.28&4.01$^{+0.48}_{-0.82}$&4.91$^{+0.59}_{-1.01}$&1.64$^{+0.19}_{-0.34}$&1.34$^{+0.16}_{-0.28}$&10.90$^{+1.30}_{-2.23}$&8.92$^{+1.06}_{-1.84}$&1.84$^{+0.26}_{-0.37}$&1.50$^{+0.21}_{-0.30}$\\
10.33&1.78&2.18&4.46/5.46&3.46&2.82&4.98$^{+0.59}_{-1.02}$&6.10$^{+0.72}_{-1.26}$&2.03$^{+0.24}_{-0.42}$&1.66$^{+0.20}_{-0.34}$&13.50$^{+1.70}_{-2.70}$&11.10$^{+1.30}_{-2.31}$&2.28$^{+0.32}_{-0.46}$&1.86$^{+0.26}_{-0.37}$\\\hline
6.71&2.80&3.43&6.99/8.56&1.43&1.17&7.81$^{+0.93}_{-1.60}$&9.57$^{+1.14}_{-1.97}$&0.84$^{+0.10}_{-0.17}$&0.69$^{+0.08}_{-0.14}$&5.61$^{+0.67}_{-1.15}$&4.58$^{+0.54}_{-0.94}$&0.94$^{+0.14}_{-0.19}$&0.77$^{+0.11}_{-0.15}$\\
10.33&3.47&4.25&8.67/10.63&1.78&1.45&9.69$^{+1.16}_{-1.99}$&11.87$^{+1.42}_{-2.44}$&1.04$^{+0.13}_{-0.21}$&0.85$^{+0.10}_{-0.18}$&6.96$^{+0.83}_{-1.43}$&5.68$^{+0.68}_{-1.17}$&1.17$^{+0.17}_{-0.24}$&0.96$^{+0.13}_{-0.20}$\\\hline
\end{tabular}
\end{center}
\end{scriptsize}
\end{table*}

\end{landscape}

Using the results from the previous section we can now calculate the upper limits on the electron densities, $n_e$, for each temperature, from the definition of the model's normalization: 
\begin{equation}\label{apec_norm}
\centering
\frac{10^{-14}}{4\pi[D_A(1+z)]^2}\int{n_e N_H dV}
\end{equation} where $D_A$ is the angular diameter distance to the source. We assumed the volume to be the one mentioned in the previous subsection, a wide spherical shell centred on the AGN and beginning just outside the edge-brightened emission around the radio bubbles. The results are displayed in Table \ref{ne_out}. Notice how the derived $n_{e,out}$ increases sharply below temperatures of 0.3 keV. This is caused by the loss of sensitivity of the instrument at low energies, and a change in the behaviour and type of emission lines found at these temperatures. We can derive from these values the pressure of the gas, also displayed in Table \ref{ne_out}. The pressures are consistent with the environment of Mrk 6 being quite cold and rarefied. Notice the increase of the pressure for very low temperatures: this is a consequence of the steep increase in the electron density estimations. Our choice of the $\sim$60 arcsec region is based on the most likely dimensions of the halo; in order to test these constraints we selected a larger region ($R_{ext}$=75 arcsec) and found that the densities and pressures change by 15\% at most.

We also estimated the limits on the electron densities for the shells, to compare them with these upper limits we obtained for the surrounding gas. This is necessary to verify that the proposed shock scenario holds, and see whether any limits on the external temperature can be derived. To do this we assumed that the dominant thermal contribution to the spectral fit we obtained for the radio bubbles (Section \ref{Bubbles}) originates in the edge-brightened, shell-like emission around them. Our extraction region covers the whole surface of the bubbles to include any photons from the nearest and farthest areas of the spherical shells, which have a lower surface brightness due to projection effects. The gas inside the bubbles is radio emitting plasma, and therefore does not contribute to our spectrum, and any emission from gas immediately outside the shells is accounted for by the background.

Using the constraints on the temperature and model normalization from our bubble fit, we then estimated the North and South shell thickness from the apparent thickness of the edge-brightened emission around the radio bubbles, which in Mrk 6 is quite evident, see Figure \ref{f01}. Since the spectral resolution is not enough to allow us to discern the detailed structure of the shells, we assumed a simple spherical geometry for each structure. From the extent of the emission we estimated for the North shell a width of $\sim$980 pc ($R_{int,N}=2.24$ kpc, $R_{ext,N}=3.22$ kpc), and for the South shell $\sim$1100 pc ($R_{int,S}=1.55$ kpc, $R_{ext,S}=2.65$ kpc). Estimating the thickness of the shells is difficult due to the poor photon statistics, hence these values were taken as a conservative upper limit. They are larger than what we expected a priori, but this may partly be due to the apparent lower density of the external environment in Mrk 6, as well as to the limited resolution we can achieve with our photon statistics. The results we obtained are shown in Table \ref{ne_shells}. This table also shows what the limits on the $n_{e}$ would be, both for the shells and the external environment, assuming a thickness of $\sim$200 pc, similar to the one estimated for Cen A.

From these values we can derive the total mass of the gas contained in each structure and the resulting pressure. These results are also displayed in Table \ref{ne_shells}. When comparing these results with NGC 3801 \citep{Croston2007}, we find that even though the electron density is lower in Mrk 6, the apparent thickness of the shells is much bigger, hence the total mass of the gas is higher by slightly less than one order of magnitude. The inferred pressures for the gas are roughly the same as the ones found on NGC 3801, due to the temperature of the gas being higher in Mrk 6. The errors on the pressure are calculated from the uncertainty in the gas temperature, 0.94$^{+0.12}_{-0.19}$ keV. We can see that choosing thinner shells, like the ones found on Cen A, would imply a much smaller ($\sim$0.5) fraction of gas, while the pressure would almost double (see lower section of Table \ref{ne_shells}). The pressure jump $P_{shell}$/$P_{out}$ for a 1 kpc shell is $\sim$7-63, higher than in NGC 3801, but again this is to be expected, due to the bigger temperature contrast and more rarefied external environment in Mrk 6. The contrast would be much higher for thinner $\sim$200 pc shells, $\sim$14-122.

We calculated the minimum internal pressure of the Northern bubble under equipartition conditions from the radio data, fitting a broken power-law electron energy spectrum with $p=2$ at low energies, steepening to $p=3$ at the electron energy that gives the best fit to the data. We assume no protons and $\gamma_{min}$=10. We obtained a value of P$\sim$4$\times10^{-13}$ Pa, consistent with the results obtained by \citet{Kharb2006}, and roughly an order of magnitude lower than the pressures we derive from our X-ray data. This departure from minimum energy is often found in FR I radio galaxies (e.g. \citeauthor{Morganti1988} \citeyear{Morganti1988}), and although there are very few examples where the minimum internal pressure is higher than the external pressure in the ISM, we also found this effect in NGC 3801 and Cen A. These results imply that there is some additional contribution to the internal pressure, caused by a large deviation from equipartition conditions, a high fraction of non-radiating particles, such as thermal or relativistic protons originated from the interaction of the galaxy's gas with the jet, or a low filling factor. Of these, the second explanation is the most plausible (see \citeauthor{Hardcastle2007} \citeyear{Hardcastle2007}, \citeauthor{Croston2008b} \citeyear{Croston2008b}), even more so if the jet is indeed precessing, as suggested by \citet{Kharb2006}, or if it is just episodic.

In the limit where there is no shock the pressures of the external medium, $P_{out}$, and the shells, $P_{shell}$ should be equal \citep{Landau6}. The difference in the apparent luminosities of both regions already rules out this situation if both temperatures are equal. We can also rule it out by noticing that these values of $n_{e,out}$ and $P_{out}$ would imply external temperatures far below 0.1 keV (see the results from Tables \ref{ne_out} and \ref{ne_shells}) which is unlikely, as discussed in the previous Section. The density ratio $\rho_{out}$/$\rho_{shell}$ tends to 4 in the case of a strong shock. For values of $\rho_{out}$/$\rho_{shell}$ smaller than 4 the required values of the external temperature are below or just above 0.1 keV; the most plausible scenario is therefore that of a strong shock.

Assuming a density contrast of 4, and using again the limits on $n_{e,out}$ from Table \ref{ne_out}, we can constrain the possible temperature for the external gas to be around 0.2 keV. This result is consistent for both the North and South shell values and for both the 1 kpc and 200 pc shell thicknesses (the 200 pc hypothesis would imply a higher gas density within the shells, thus to maintain the density contrast of 4 the upper cap on the external temperature would be lowered, but the results still imply $kT>0.1$ keV). We were able to directly measure the temperature of the external medium in Cen A \citep{Kraft2003}, and found the temperature of the ISM to be $\sim$0.29 keV. For NGC 3801 \citep{Croston2007} we obtained a value of 0.23$^{+0.21}_{-0.09}$ keV. These values directly applied to Mrk 6 would cause an inconsistency, since the density ratio cannot exceed a value of 4, but given the uncertainties they are still in very good agreement with the current results.

As an additional test we repeated our calculations for Z=0.6$\odot$ and Z=1.0$\odot$. The model normalizations, and consequently the electron densities, inversely depend on the abundance, so we found lower values as we approached the solar values. However, the impact of this effect is bigger on the results from the external gas than to that contained in the shells. This causes an inconsistency when applying the conditions for a strong shock, since the external gas would be required to have temperatures far below 0.1 keV, which are implausible, as we discussed above. Higher abundances also have an impact on the energy yield. The pressure and total energy for Z=0.6$\odot$ would be $\sim20\%$ lower than the ones we calculated with Z=0.35$\odot$, and $\sim40\%$ lower for Z=1.0$\odot$.

To further test these results, we used the B-band luminosity of Mrk 6 to relate it to other galaxies following the relations from \citet{Osullivan2001,Osullivan2003}. We found that both the temperature of the external gas (0.2 keV) and the abundance we fit to the data (0.35$\odot$) are consistent with their relations. Their sample includes two galaxies with gas X-ray and B-band luminosities slightly larger ($L_{X}\sim10^{40}$ erg s$^{-1}$, $L_{B}\sim10^{11}L_{\odot}$) than what we found for Mrk 6 where these parameters can be directly related. For NGC 1549 they find an outer gas temperature $\sim$0.25 keV and an abundance $\sim$0.14$\odot$. For NGC 4697 their results show T $\sim$0.24 keV and Z $\sim$0.4$\odot$. We therefore consider our best estimates of the properties of the large-scale hot gas in Mrk 6 to be plausible ones. 

We can calculate the Mach number for the shock, as follows \citep{Landau6}:

\begin{equation}\label{Mach}
\centering
\mathcal{M}=\sqrt{\frac{4(\Gamma+1)(T_{shell}/T_{out})+(\Gamma-1)}{2\Gamma}}
\end{equation} where $\Gamma$ is the polytropic index of the gas, assumed to be 5/3. The results for our chosen temperature range are displayed in Table \ref{Mtable}. 

\begin{table}[h]
\begin{scriptsize}
\begin{center}
\caption{Possible values of the Mach number given by our temperature constraints. The values of $T_{shell}$ are the best fit value and the 1$\sigma$limits allowed by the errors. The most likely value is shown in bold.}\label{Mtable}
\setlength{\extrarowheight}{2pt}
	\begin{tabular}{c|ccc}\hline\hline
    &\multicolumn{3}{c}{$kT_{shell}$ (keV)}\\
    {$kT_{out}$ (keV)}&0.75&0.94&1.06\\\hline
    0.10&4.92&5.50&5.84\\
    0.20&3.49&\bf{3.90}&4.14\\
    0.30&2.86&3.20&3.39\\
    0.40&2.49&2.78&2.95\\
    0.50&2.24&2.49&2.64\\
    0.60&2.05&2.28&2.42\\\hline
	\end{tabular}
\end{center}
\end{scriptsize}
\end{table}

From the previous constraints we can assume the most likely value for the temperature to be $0.2^{+0.1}_{-0.1}$ keV. This yields a value for the Mach number of  $3.90^{+1.94}_{-1.04}$, of the same order of magnitude as the one we obtained for NGC 3801 \citep{Croston2007}, although smaller than the one found for Cen A \citep{Kraft2003}. This is to be expected, since the morphology of Mrk 6 is more similar to the one of NGC 3801 than to that of Cen A, where only one shell is detected.

The speed of sound in the ISM ranges from $\sim$190 km s$^{-1}$ in the case where the external temperature is 0.1 keV to $\sim$460 km s$^{-1}$ when $kT=0.6$ keV. Since the Mach number also varies with the temperature (see Table \ref{Mtable}), we derived a consistent value of 1030$^{+75}_{-110}$ km s$^{-1}$ for the speed of the gas. This allows us to derive the kinetic energy of the gas, which is displayed in Table \ref{ne_shells}. The total thermal energy and total work available for each radio bubble are also displayed in Table \ref{ne_shells}.

The total energy (thermal + kinetic) from both radio bubbles, under the 1 kpc width assumption, is $\sim2.6-4.6\times10^{56}$ erg. This corresponds to $\sim$3.6$\times 10^{5}$ supernova explosions with individual explosive energy of $10^{51}$ erg. This result also agrees with the value derived by \citet{Kharb2006} from the radio data of Mrk 6, $\sim1\times10^{55}$ erg, since it was a lower limit. They estimate a kinetic luminosity for the jet of $\sim10^{42}$ erg s$^{-1}$, derived from an inferred radio luminosity of $\sim10^{40}$ erg s$^{-1}$ and an efficiency of 1\%. With our results, the jet would require $\sim1.1\times10^{7}$ years to deposit all its energy into the nuclear ISM and produce this pair of radio bubbles. This result is almost two orders of magnitude larger than the one they obtained, but not incongruent, since their estimation was a lower limit derived from the equipartition pressure. These results also agree with the timescales they propose in the context of a precessing jet. Another timescale can be inferred from the supersonic expansion speed of the bubbles, which would require a time of $\sim 2.5-3.1 \times 10^{6}$ years to inflate to their current size, assuming $M \sim 4$ throughout their lifetime. If we adopt this timescale, the required jet power is $\sim 7 \times 10^{42}$ erg s$^{-1}$.

The thermal energy stored in the shells is an order of magnitude larger than the kinetic energy required to inflate the cavities (see Table \ref{ne_shells}), meaning shock heating must be the dominant energy transfer mechanism during this stage of the evolution of the radio source. Although the emission from the surrounding gas is too weak for us to fit a model to its brightness profile, which would be ideal to determine the direct impact of this thermal energy into the surrounding ISM, our results, and comparison with those obtained for Cen A and NGC 3801, show that this energy is enough to significantly alter the surrounding ISM, sweeping out the remaining halo gas.

With our current photon statistics we have no direct evidence of changes in the brightness (and therefore gas density) across the shells, therefore we cannot state whether these structures are brighter in the inner edge, closer to the nucleus, in Mrk 6. However, this would be expected if the bubble expansion is supersonic everywhere: even though the shock is weakest close to the nucleus, the ISM is much denser in this direction, hence the associated X-ray emission is higher.

If we assumed the density of the gas not to be constant across the shell, with a higher density in the area closer to the radio source, the derived constraints on the temperature of the external medium would be relaxed, allowing higher values. Consequently, the pressure and density of the external gas could also be higher. This situation would bring our results closer to the values we obtained for Cen A and NGC 3801. While this would be desirable for consistency, we must note the scenario in Mrk 6 is quite different, both due to its morphology and to its lower radio power, thus we could indeed be seeing a case of the AGN driving a strong shock into a very cold environment. It could also be the case that the shells are not continuous all around the radio bubbles, which would change the constraints on the density of the gas. This hypothesis could be tested with a longer exposure.

To conclude our analysis and study the fueling mechanism of the AGN, we used our estimates on the external gas temperature and density to calculate the Bondi accretion rate from the central hot gas. We used the relations from \citet{Marconi2003} and the K-band luminosity of Mrk 6 and obtained a mass for the black hole of 3.7$\times 10^{8} M_{\odot}$. The Bondi rate is given by
\begin{equation}\label{Bondi}
\centering
\dot M=4 \pi \lambda \rho G^{2} M_{BH}^{2} c_{S}^{-3}
\end{equation} where $\lambda$=0.25, $\rho$ is the density of the gas at the Bondi radius, and $c_{S}$ is the speed of sound in the medium ($c_{S}=\sqrt{\Gamma kT/ \mu m_p}$). Since the density and temperature at the Bondi radius cannot be lower than the ones we estimated for the gas outside the shells, we used those values to obtain a lower limit of $\dot M=2.04 \times 10^{-5} $M$_{\odot}$ yr$^{-1}$. Assuming $\mu$=0.1 this translates to a Bondi power $P_{B}=1.1 \times 10^{41}$ erg s$^{-1}$. The energy output we measured from the AGN bolometric luminosity ($\sim 2.4 \times 10^{44}$ erg s$^{-1}$, using a bolometric correction factor of 19) and the shocked gas shells ($\sim 7 \times 10^{42}$ erg s$^{-1}$) imply an accretion rate of $\dot M\sim5.7 \times 10^{-2}$ M$_{\odot}$ yr$^{-1}$. This suggests that Bondi accretion is unlikely to provide all the energy we observed to be transferred to the ISM, even with an efficiency of 100$\%$. Assuming $\mu$=0.1, the density would have to be over three orders of magnitude higher than our estimates of the density of the unshocked gas in order for Bondi accretion to be viable. If Bondi accretion cannot account for the energy we are observing, the main fuel source for this system is likely to be cold gas. Cold accretion does not depend on the black hole mass and can happen in poor environments as long as there is a fuel reservoir, which may be the case in Mrk 6 if the central obscuring material seen in the optical is indeed cold gas from a past merger.
%
\section{Conclusions}\label{conclusions}

We conclude that there is a high probability that we are observing shells of strongly shocked gas around both radio bubbles of Mrk 6, with Mach numbers 3.2-5.5 consistent with the Rankine-Hugoniot  conditions for a strong shock. This is the third clear detection of such a process in a low-power radio source, and therefore reinforces the hypothesis that this mechanism may be very common and may play a fundamental role in the process by which these young radio sources form their stellar populations.

It is very likely that much of the gas in the shells will escape the gravitational potential of the host galaxy. Our results show shock heating is the dominant process, thermal energy being an order of magnitude larger than the work required to inflate the radio cavities, implying the outburst impact on the host galaxy's ISM is likely to be dramatic. Moreover, the inner set of bubbles is expanding perpendicularly to the external ones, and into denser regions of the host galaxy. Although the shock will probably be weaker, due to the increased density, the expansion of the inner bubbles is likely to cause a more direct disruption on the ISM and are likely to have a bigger impact on star formation. Unfortunately the resolution of the current generation of X-ray telescopes does not allow us to determine the details of this ejection of gas.

We also conclude that the hypothesis of a variable absorption gas column, with timescales of 2-6 years, caused by clumpiness of the accreting gas close to the black hole, is a likely explanation for the variations we see between the \textit{Chandra} and \textit{XMM} spectra. If this is indeed the case, and as already pointed out by \citet{Risaliti2002}, this scenario could be more common than previously thought, and applicable also to intermediate Seyfert systems, such as Mrk 6. Observations of more such sources may lead to better constraints on the geometry and composition of the gas distribution surrounding the AGN. Further observations at soft and hard energies, and detailed X-ray spectroscopy of this source would be of extremely useful for this purpose.

The variable obscuration also poses some questions about the mechanism fueling the AGN. Our analysis of the AGN core shows that the dominant contribution to the soft excess in this system comes from the AGN non-thermal emission, although there is some contribution from the jet, which could explain the flat power law spectra we are observing. Our results on the energetics of the source show that Bondi accretion might not be enough to power it, so that it is possible that cold gas accretion is the dominant mechanism in this system, unlike what is seen in many more powerful radio sources \citep{Hardcastle2007,Balmaverde2008}. Cold gas has also been suggested as the primary source of fuel for Cen A and NGC 3801 \citep{Croston2007}. It is also possible that the active nucleus of Mrk 6 is accreting both the hot phase of the ISM and cold gas from a past merger. This is consistent with recent results on early-type galaxies by \citet{Pellegrini2010}.

While the precessing jet scenario is challenging, due to the high torque required to rotate the axis of the black hole nearly 180$^{\circ}$ within a time of $\sim 10^{6}$ years, required to create both perpendicular sets of bubbles, it cannot be ruled out. The dust lane hinted in the optical images could be a sign of a merger, which could have provided the necessary torque, and the evidence for obscuration variability near the black hole could be further evidence. It would be interesting to follow this hypothesis and find any correlations with other systems where a similar variation in absorption has been studied. However, this evidence can also be interpreted in terms of a cluttered and uneven environment, which can be inferred on bigger scales from the [OIII] emission distribution, causing the radio bubbles to expand asymmetrically and in different directions. Again, more evidence, both from future observations of Mrk 6 and from other sources, will be useful to achieve an accurate interpretation of the underlying physics behind the variable absorption and the asymmetries we have observed.

\subsection*{Acknowledgements}

BM thanks the University of Hertfordshire for a PhD studentship. MJH thanks the Royal Society for a research fellowship. JHC acknowledges support from the South-East Physics Network (SEPNet). This work was partially supported by NASA grant GO9-0117X. This work is based on observations obtained with \textit{XMM-Newton}, an ESA science mission with instruments and contributions directly funded by ESA Member States and NASA. It has also made use of new data from \textit{Chandra} and software provided by the Chandra X-ray Center (CXC) in the application package CIAO. We thank the anonymous referee for the useful comments.

\bibliographystyle{apj}
\bibliography{Mrk6}

\end{document}